\documentclass[11pt]{article}

\usepackage[utf8]{inputenc}
\usepackage[T1]{fontenc}

\usepackage[a4paper,margin=2.5cm]{geometry}
\usepackage{amsmath}
\usepackage{amssymb}

\usepackage{booktabs}
\usepackage{array}
\usepackage{tabularx}
\usepackage{ragged2e}
\newcolumntype{L}[1]{>{\RaggedRight\arraybackslash}p{#1}}

\usepackage{graphicx}
\usepackage{caption}
\usepackage{listings}
\usepackage{xcolor}
\definecolor{codebg}{gray}{0.96}
\usepackage[hidelinks,hyperfootnotes=false]{hyperref}

\usepackage[hang,flushmargin]{footmisc}

\title{\textbf{Canonicalization Failures as a\\ Recurring Vulnerability Class:\\ Representation Divergence in Cryptographic Systems and Its Avoidance}}
\author{Arslan Br\"omme\thanks{Dipl.-Inform., B.Sc., CISSP, CISA, CISM, CAISE. \texttt{arslanb@chain-horizon.com}}\\ Chain Horizon GmbH\\[0.3em] Version 0.9.4.4 (Working Draft)}
\date{}

\begin{document}
\maketitle
\thispagestyle{empty}

\begin{center}
\fbox{\parbox{0.9\textwidth}{\itshape\small Preprint / working paper. This version is a work in progress and may be updated. Comments are welcome. This document is an English translation of the corresponding German version (v0.9.4.2) of this paper.}}
\end{center}

\vspace{1em}

\section*{Abstract}

Cryptographic systems operate on bytes but mean semantic objects. The translation between the two is rarely unique. Where this uniqueness is not enforced, an attack surface opens up as soon as a hash, a signature, replay protection, or consensus identity depends on the representation. The same class of failure has been discovered independently and named locally across many ecosystems, as transaction malleability, non-deterministic value encoding, message malleability, or hash chain malleability, without its common root being tracked as a cross-ecosystem grid. This work organizes the scattered findings systematically: it shows that they are instances of \emph{one} violated uniqueness condition, along two basic directions: multiple valid codes for one object (object-side multiple representation) or one code for multiple objects (code-side semantic collapse). The contribution is explicitly not the discovery of the phenomenon, but is threefold: the systematization by representation mechanism rather than by affected system, the bridge between classical canonicalization security and a representation- and computability-theoretic foundation, and the translation into an applicable review procedure (a canonicalization obligation with a sequence of review steps, a field-type classification, and an operational boundary model) with which the risk can be recognized preventively. We substantiate the class with worked-out cases, delimit it against incidents that are \emph{not} representation problems, and deliberately keep the claims to what is demonstrable: enforced uniqueness can reduce the exploitability of a failure class, but it neither replaces further protective measures nor makes any statement about cryptographic security in the narrower sense.

\vspace{0.5em}
\noindent\emph{Keywords}: canonical serialization, representation theory, malleability, signature uniqueness, blockchain security, byte-input invariance, canonicalization attack

\newpage

\section{Introduction}

Cryptographic systems process finite byte sequences, but they \emph{mean} semantic objects: a number, a signature, a state. Between the two lies a translation, the representation, and this translation is rarely unique. The same semantic object can occur in several admissible byte forms, and conversely several meanings can collapse onto the same form. As long as a system does not deliberately restrict this multiplicity, a gap opens up between what a value \emph{means} and what a system \emph{sees}.

This gap has consequences, because central cryptographic operations (hashing, signing, comparing) do not work on the meaning but on the bytes. When two forms mean the same thing but are different bytes, an attacker can play one off against the other. When two meanings yield the same bytes, an attacker can pass one off as the other. At this seam, expensive failures have arisen repeatedly in practice: from the signature that exists in two valid forms, through the transaction hash that can be altered unnoticed, to the storage value where ``empty'' and ``confirmed'' become indistinguishable. What is remarkable about these failures is not that they occur individually, but that in almost every cryptographic ecosystem they are \emph{independently rediscovered} and there tracked \emph{under their own local name}. Bitcoin knows ``transaction malleability'', Cosmos ``non-deterministic value encoding'', chain-agnostic profiles ``message malleability'', zero-knowledge circuits the non-canonical field-element representation (``aliasing''), and the software supply chain, recently, ``hash chain malleability''. Each of these fields has named the same basic pattern for itself, without this root being tracked in the blockchain and crypto-protocol security literature as a common, cross-ecosystem grid.

This root is not a cryptographic one but a representation-theoretic one. The accompanying foundational paper [1] makes one property the central requirement: the uniqueness condition, according to which exactly one valid code belongs to each semantic object, formally $\mathrm{dec}^{-1}(\{x\}) = \{\mathrm{enc}(x)\}$. Where this condition is satisfied, the representation gap described above is structurally closed. Where it is violated, an attack surface opens up as soon as security decisions depend on the ambiguous representation. And it can be violated in only two ways: on the one hand several valid codes carry the same object, the direction known as malleability; on the other hand one code carries several meanings, the direction of collision and of sentinel collapse. These two directions are the organizing principle of the present work. Before we name the contribution, we delimit it against an obvious objection, since the underlying security principle is not new. Classical software and operating-system security has long known that multiple representations of the same object can subvert a check, for example under names such as ``canonicalization'' attacks, in the resolution of equivalent file paths, or in authorization via non-canonical URLs [2]. Likewise, the practical necessity of deterministic and canonical serialization is documented in detail in format standards [3, 4]. What these established works lack is twofold: the connection of the security phenomenon with its \emph{mathematical} foundation, representation and computability theory, and its systematic application to the specific mechanisms of cryptographic and distributed systems, in which an encoding difference not only subverts a condition but strikes the identity of a hash, signature, or consensus object.

This is precisely where the contribution of this work lies. It is not the discovery of a new phenomenon (which would be untenable given the precursors named above), but is threefold. First, the \textbf{systematization}: we order the scattered vulnerabilities by their representation mechanism rather than by affected system, and show that they are instances of \emph{one} class with \emph{two} break directions. Second, the \textbf{foundational bridge}: we connect the classical canonicalization security principle with the representation-theoretic foundation of the accompanying paper, thereby making visible why the same failure class reappears in every ecosystem, because its common, computability-theoretic nature has so far not been tracked as such a grid in the blockchain and crypto-protocol security literature. Third, the \textbf{transfer into practice}: we illustrate the class with cases worked out in detail, name an instance not previously tracked as a class of its own in the Ethereum context, and derive from it a canonicalization obligation together with a sequence of review steps for developers and auditors, which we condense into a defensively framed guide.

We deliberately keep the claims to what can be clearly defended. As an overarching statement across all cases, the weaker and demonstrable claim holds throughout (a canonical representation could have \emph{reduced the exploitability of a failure class}), while the stronger claim, that a failure class is structurally \emph{excluded}, appears only where the individual case supports it. Real security incidents are almost always multi-causal. We therefore nowhere claim that canonicalization alone would have prevented a concrete loss. Where the representation-theoretic perspective reaches a limit or an incident has a different cause, we say so explicitly.

The work understands itself as a continuation of the program of a protocol-relative representation perspective, which the foundational paper [1, Section 11] explicitly announced as follow-up work. It introduces no new cryptographic primitive and proves no new impossibility result. Its contribution lies in systematization, transfer, and the connection of two hitherto separate perspectives. It makes no statement about cryptographic security in the sense of secrecy or collision resistance. Canonicity is a property of the representation, not a security guarantee.

The remaining structure follows this arc. Section~2 compactly summarizes the required representation principle and its two break directions. Section~3 introduces the unified analysis scheme with which the cases worked out in detail are examined in Section~4 and contrasted with their canonicalization-guided versions in Section~5. Section~6 spans the full grid, places the cross-ecosystem evidence, acknowledges the precursors, and works out the paper's own contribution as a bridge. Section~7 derives from it an applicable review procedure: a canonicalization obligation, a decomposition of composite objects, a field-type classification, and an operational boundary model. Section~8 delimits where representation divergence is \emph{not} the cause; Sections~9 and~10 discuss limits and consequences. An appendix translates the insights into defensively framed corrections and a developer checklist, with executable, tested Python artifacts and compilable Solidity versions.

\section{The Representation Principle}

This work builds on a single idea from the foundational paper [1], which we summarize here compactly enough that the remaining text stays readable without having read it: the distinction between the \emph{value} of a semantic object and its \emph{presentation} as a concrete finite description. One and the same value (for instance the amount one hundred, or a particular digital signature) can be described by different byte sequences. Whether a system behaves correctly often depends not on the value, but on how uniquely the assignment between value and byte sequence is regulated.

\subsection{The Uniqueness Condition}

At the core lies a notion that the foundational paper introduces as \emph{canonical encoding} [1, Definition 3]. Intuitively it requires: to each semantic object belongs exactly one valid byte sequence. More precisely, one considers two mappings---an encoder $\mathrm{enc}$, which assigns a byte sequence to an object, and a decoder $\mathrm{dec}$, which reads a byte sequence as an object. Canonicity is achieved when, for each value $x$, the set of all valid codes that $\mathrm{dec}$ decodes to $x$ (``that $\mathrm{dec}$ reads as $x$'') contains exactly one element:
\[ \mathrm{dec}^{-1}(\{x\}) = \{\mathrm{enc}(x)\} \]
We refer to this in the following, for short, as the \emph{uniqueness condition}. The notation $\mathrm{dec}^{-1}(\{x\})$ here means the \emph{preimage} of $x$ under the decoder, that is, the set of all codes $c$ with $\mathrm{dec}(c) = x$. It does not presuppose that $\mathrm{dec}$ is invertible; rather, the uniqueness condition requires precisely that this preimage be a singleton and coincide with the representation chosen by $\mathrm{enc}$. It is stronger than mere reproducibility: a procedure can reliably produce the same output for a fixed input (determinism) and still admit different inputs that mean the same thing. The foundational paper separates these two properties explicitly [1, Section 3.4]: an encoder that always outputs ``1/2'' is deterministic. But if its decoder is willing to read ``2/4'' as well as the same value, then the encoding is not canonical. The uniqueness condition excludes precisely such second representations.

For the operational side, the foundational paper coins the notion of \emph{byte-input invariance} [1, Section 1]: the requirement that semantically equal presentations be mapped to the same operational byte sequence. Cryptographic operations need exactly this property when they are to represent semantic equality through equality of their byte inputs, for instance so that two descriptions of the same object produce the same hash or the same signature input. The vulnerabilities considered in this work are, in this language, operational violations of byte-input invariance.

\subsection{The Two Break Directions}

The uniqueness condition is a statement about a mapping between two sets, the semantic objects and the valid codes. Such an assignment can deviate from uniqueness in exactly two ways, and these two ways form the organizing principle of the entire work.

The first direction violates the \textbf{uniqueness of codes per object}: several valid byte sequences denote the same semantic object. This is the situation known in cryptography under the name \emph{malleability}: an object exists in several equally accepted forms. Formally, the set $\mathrm{dec}^{-1}(\{x\})$ is larger than a singleton. There exist codes $c_1 \neq c_2$ with $\mathrm{dec}(c_1) = \mathrm{dec}(c_2) = x$. Whoever sees one of these forms can produce the other and play it off against it. Viewed from the encoding side, this means: the encoder $\mathrm{enc}$, which assigns to each object its canonical form, does not produce these second representations, yet the decoder $\mathrm{dec}$ nevertheless assigns them to the same object. They are superfluous representations of already canonically encoded objects, not codes without an object. For $\mathrm{enc}$ to be surjective would mean precisely that every accepted code is produced by $\mathrm{enc}$; the surjectivity break consists in there being accepted codes beyond that. The second direction violates the \textbf{uniqueness of the object per code}: a single code is shared by different semantic objects. This is the situation of \emph{collision} and, in its form typical of state systems, of \emph{sentinel collapse}: an absence or default value coincides with a valid value. Here, distinct semantic objects share the same accepted code, so the system can no longer distinguish them.

For naming, we pair a descriptive and a mathematical name in each case and use them consistently in the following: the first direction we call \textbf{object-side multiple representation}---a \emph{surjectivity break} of the encoding mapping $\mathrm{enc}\colon \text{object} \to \text{code}$, in short \emph{S-break}: multiple codes for one object. The second direction we call \textbf{code-side semantic collapse}---an \emph{injectivity break} of the same mapping, in short \emph{I-break}: one code for multiple objects. The mathematical names refer to the direction of the representation condition chosen in the foundational paper, from the semantic objects to the codes, not to the naive injectivity of a decoder alone. The uniqueness condition $\mathrm{dec}^{-1}(\{x\}) = \{\mathrm{enc}(x)\}$ from the foundational paper [1, Definition 3] requires precisely that $\mathrm{enc}$ establish a bijective correspondence between the semantic objects and the valid codes. Its two directions yield the two names: an accepted code that does not lie in the image of $\mathrm{enc}$ is a superfluous second representation of an already canonically encoded object, and thus a break of the surjectivity of $\mathrm{enc}$, while two objects with the same code violate the injectivity of $\mathrm{enc}$. Surjectivity here is not meant as a statement about an arbitrary set of codes, but relative to the valid code range accepted by the system: if the system accepts more codes than $\mathrm{enc}$ produces as canonical representations, then the additional ones lie outside the image of $\mathrm{enc}$. The defect in the S-break thus lies in the superfluous accepted representation, not in $\mathrm{enc}$ itself producing multiple codes for one object. Both names thus denote the same loss of one half of the bijection, in each case seen from the encoding side: the S-break the failure of surjectivity, the I-break that of injectivity. In the explanatory text we predominantly use the descriptive names; in tables and terse additions, the mathematical short forms. The naming is not an end in itself: it allows seemingly unrelated security failures to be ordered by which of the two directions they violate, independently of the system in which or the technical level at which they occur. It is precisely this ordering that the following sections develop. This division into two is complete, provided one considers it relative to a fixed set of valid codes and a fixed semantic equality relation: if exactly one code is assigned to each object and exactly one object to each code, then the assignment is a bijection between the semantic objects and the valid codes. A mapping that breaks in \emph{neither} of the two directions thereby satisfies the uniqueness condition. Relative to this fixed basis, the two break directions exhaust the violations of the uniqueness condition. We thereby deliberately exclude questions that lie prior to this basis (such as faulty or non-terminating decoders, invalid codes, or an unfixed equality relation). They concern the well-definedness of the assignment itself, not the direction of its break. The two directions are at the same time independent: a system can break in one and be intact in the other, as the case studies will show.

\section{Analysis Scheme}

The case studies of the following sections could easily fall apart into a collection of individual stories: each incident with its own backstory, its own technical detail, its own lesson. To avoid this and to keep the common structure visible, we subject each case to the same four-step grid. It is introduced here once and merely applied in Sections~4 and~5. The repetition of the form is intentional, for it is what turns individual cases into a class.

For each incident considered, we ask four questions in a fixed order:

\textbf{First---mechanism.} What exactly goes wrong at the byte or state level? We describe the technical core as briefly as possible and as precisely as necessary, without the surrounding backstory of the incident.

\textbf{Second---classification.} Which of the two break directions is present, an object-side multiple representation (surjectivity break, S-break: multiple codes for one object) or a code-side semantic collapse (injectivity break, I-break: one code for multiple objects)? And at which level does uniqueness break: at the value itself, at the composition of structured data, or in the state and storage model? This classification is the step that locates the case within the common framework.

\textbf{Third---effect of canonicalization.} Would an enforced uniqueness condition have reduced the exploitability of this failure class? Here we observe the terminological convention fixed in the introduction: the overarching statement is that canonicalization could \emph{reduce} exploitability. The stronger statement, that it structurally \emph{excludes} the failure class, we use only where the individual case supports it.

\textbf{Fourth---what additionally remained necessary.} Real incidents are multi-causal. We therefore state explicitly which further measures, beyond canonicalization, would have been necessary, so that the case is not misunderstood as evidence for an overstated thesis. This step is the honest counterweight to the third: it records that an enforced uniqueness condition acts structurally within the respective isolated representation-failure class, but does not prevent the entire incident: it is effective within its class, not a sufficient protection against all contributing causes. Where other measures could likewise have averted the same concrete loss, we say so.

The grid accomplishes two things at once. It disciplines the presentation by reducing each case to the same core and keeping narrative embellishment at bay. And it makes the central thesis testable: when different incidents fall into the same few categories at step two, this is not a rhetorical device but the actual finding that we are dealing with a class and not a list. A case that eludes the grid is thereby not a nuisance but instructive: Section~4.3 deliberately presents such a case in order to mark the boundary of the framework.

\section{Case Studies}

We apply the analysis scheme to three incidents. The first two are instances of the vulnerability class, one per break direction: ECDSA malleability as a surjectivity break at the value level, the Nomad incident as an injectivity break at the state level. The third, the Wormhole incident, is deliberately \emph{not} an instance---it marks the boundary of the framework and guards against the temptation to reinterpret every expensive security breach after the fact as a representation problem.

\subsection{ECDSA Signature Malleability}

\textbf{Mechanism.} An ECDSA signature consists of a pair of numbers $(r, s)$. The check of whether a signature is valid uses $s$ only in the form of its multiplicative inverse. This has an unintended consequence: if one replaces $s$ by $n - s$ (the $s$ ``mirrored'' at the curve order $n$), the signature remains valid. For every valid signature $(r, s)$ there thus exists a second, equally valid signature $(r, n - s)$ for the same message under the same key. The same signature is present in two admissible forms.\footnote{Formally: in the verification equation, $s$ enters only via $s^{-1}$. Since $(n - s)^{-1} = -s^{-1} \bmod n$, the curve point computed during verification is mirrored. Its $x$-coordinate (and thus the comparison value $r$) remains unchanged. With $(r, s)$, therefore, $(r, n - s)$ also satisfies the same equation. The explanation refers to the ordinary ECDSA verification of a pair $(r, s)$; in contexts with an additional recovery identifier (such as $v, r, s$ in Ethereum signature recovery), this must be carried along separately. This inherent malleability is to be distinguished from encoding malleability, in which the same signature takes on several forms through differing byte encoding of its elements. The latter is addressed in Bitcoin by strict DER encoding (BIP-66, status \emph{Deployed}), which however only fixes the \emph{form} of the elements, not their \emph{values}.}

\textbf{Classification.} This is an object-side multiple representation (S-break) at the value level: two different valid codes denote the same semantic object, the signature. In the language of Section~2, the set $\mathrm{dec}^{-1}(\{x\})$ for the signature $x$ is not a singleton but contains at least the two codes $(r, s)$ and $(r, n - s)$. The uniqueness condition is violated.

\textbf{Effect of canonicalization.} Here the stronger statement is justified: an enforced uniqueness condition structurally excludes this failure class. One achieves it by admitting exactly one of the two mirrored halves as valid: the common rule restricts $s$ to at most half the curve order, $s \leq n/2$, the so-called low-S normal form.\footnote{Careful attention to normative anchoring is warranted here: in Bitcoin, BIP-62 and BIP-146 were not rolled out as consensus rules. Both are today tracked in the BIP repository with status \emph{Closed}. BIP-66 is \emph{Deployed} and concerns strict DER encoding, not the low-S value normal form. Reference implementations such as libsecp256k1 produce signatures with low $s$, without this being an enforced Bitcoin consensus rule. In Ethereum, by contrast, the low-S rule is normatively anchored: EIP-2 declares transaction signatures with $s > n/2$ invalid and justifies this explicitly with malleability: $s$ can be mirrored to $n - s$ and $v$ flipped along with it, without losing validity [22]. Decisive for Section~6.4 and the appendix is the restriction likewise recorded there: the precompile \texttt{ecrecover} remains unchanged and continues to accept high $s$-values. Whoever uses it in a contract must therefore enforce the canonical signature form themselves. The consensus rule does not cover them.} With this, exactly one valid code belongs to each signature. The second is inadmissible by encoding rule. The malleability disappears not through a cryptographic strengthening, but through a representation decision: it is the same mathematical object, only with enforced uniqueness of form.

\textbf{What additionally remained necessary.} The low-S rule removes one of several malleability sources: the one from the mirrored $s$. It makes no statement about the security of the signature scheme itself and does not remove other ways of reshaping a transaction. In Bitcoin, for instance, the malleability consequential for transaction identity became manageable only with a structural measure that keeps the signature data out of the computation of the transaction identifier.\footnote{SegWit (BIP-141) separates the witness data from the computation of the transaction ID, so that the txid is stable from signing onward. Activation took place on 24 August 2017 (UTC) at block height 481{,}824 [5].} The representation correction is necessary here, but it is part of a larger picture. The complete removal of the practical consequences required an additional, structural step.

\subsection{Nomad --- Collapse of Absence and Validity}

\textbf{Mechanism.} The Nomad bridge checked whether an incoming message had already been confirmed as valid, via a mapping from Merkle roots to confirmation times (\texttt{confirmAt}). During an update of the contract logic, this mapping was initialized such that the value \texttt{bytes32(0)} was assigned the confirmation time 1. This is consequential, because the EVM storage model returns no error for an unset entry, but the default value \texttt{bytes32(0)}. Concretely, two mapping stages are involved: the first, from the message to its confirmed root, returns for an unknown message the default value \texttt{bytes32(0)}. The second, from the root to its confirmation time, returns for exactly this value, because of the initialization, a valid time. An unknown (also a forged) message thus passed through both stages and was ultimately considered confirmed. The responsible checking function thereby returned ``valid'' for every unknown message. Arbitrary messages could be processed without prior proof.\footnote{Verified against the post-mortem of Nomad as well as independent analyses (including Immunefi, Halborn, CertiK, Coinbase). The vulnerable behavior of the processing logic was introduced by an upgrade on 21 June 2022. Active exploitation began on 1 August 2022 and led to a loss of over 190 million USD. Since processing required no valid signature of an off-chain attester, the watchers provided for that purpose did not trigger. Characteristic, moreover, was that after the first successful transaction, arbitrary third parties could merely swap the recipient address in the message and repeat the attack, from which a massively imitated exploit arose.}

\textbf{Classification.} This is a code-side semantic collapse (I-break) at the state level: the one code \texttt{bytes32(0)} carries two semantically different meanings: ``no root set / unknown'' and ``a concrete state marked as confirmed''. The decoder, here the checking logic, can no longer tell the two apart. The absence case and a valid case share the same code. In terms of the framework, the separation between the absence or default value and the range of valid values is missing: a missing sentinel disjoint from the value range. Characteristic of the state level is that the collapse arises not from an ambiguous encoding of a value, but from the default semantics of storage, which returns an unset entry as the same zero value as a deliberately set one.

\textbf{Effect of canonicalization.} Here too the case carries the stronger statement, though with an important restriction in the next step: an enforced separation of absence and validity would have eliminated this concrete failure class. If one requires that the absence code can never denote a valid state (for instance through an explicit confirmation marker independent of a default value, which is excluded for the zero value by invariant), then the collapse between ``unknown'' and ``confirmed'' cannot arise. The failure lies not in a faulty computation, but in the default semantics of the storage model. The correction is a representation decision about the meaning of the zero value.

\textbf{What additionally remained necessary.} The case shows concretely that canonicalization, while it excludes the failure class, does not prevent the incident on its own. A preceding security audit had already noted a closely related problem---an empty message that could falsely count as already proven---as a finding. Its significance was, however, underestimated.\footnote{The audit finding QSP-19 ``Proving With An Empty Leaf'' (Quantstamp, report of 9 June 2022, classified as \emph{Low Risk}) named the possibility that an empty leaf message is falsely marked as proven [6]. To the Nomad team's objection that it is practically impossible to find the preimage of the empty leaf, the auditors replied that the problem concerned not the preimage but the provability of empty bytes as contained in the tree: the significance was thus recognized, but underestimated.} For this class, sentinel separation is the structural countermeasure, but it replaces neither invariant tests nor the appropriate response to audit findings. The framework names precisely which failure class struck here and how it could have been structurally avoided: no more, but also no less.

\subsection{Wormhole --- The Boundary of the Framework}

The Wormhole incident is presented here not as an instance of the vulnerability class, but for delimitation. It shows that not every expensive bridge collapse is a representation problem, and thereby guards the framework against the reproach that it reinterprets every security breach after the fact as a canonicity failure.

\textbf{Mechanism and classification.} The signature check of the Wormhole bridge on Solana confirmed that a cryptographic pre-check had taken place, by reading out a particular system state. It used for this an outdated function that did not check whether the passed account was actually the genuine system account. An attacker submitted a forged account. The check read the manipulated data and considered the signatures verified. Through the confirmation thus obtained by deception, the issuance of 120{,}000 wrapped Ether was authorized.\footnote{Verified against concurring analyses (including Halborn, CertiK, Kudelski, Sec3). The incident of 2 February 2022 concerned the Solana side of the bridge. Loss around 326 million USD. Remarkable for the classification as a non-representation failure: a fix using the validating variant of the function was already committed in the public repository, but not yet rolled out to the mainnet.} The cause is a missing validation of the identity of an input account, tracked in the literature as \emph{account confusion}, not an ambiguity of the representation. There is here no second valid code for the same object and no collapse of two meanings onto one code. The uniqueness condition is not touched at all. The failure lies in the fact that a check that should have taken place was omitted. The forged account is thereby not a second valid representation of the genuine system account, but an invalid input that was accepted for lack of an authenticity check. The difference from representation divergence is precisely this: there, both forms are valid; here, one is simply not checked. The case thus belongs to a different class, and that is the point: an enforced canonicalization would not have prevented it, because the problem lies not in the assignment of objects to codes, but in an omitted authenticity check. Whoever takes the present framework seriously must explicitly exclude such cases: not because they were unimportant, but because their conflation with representation failures would dilute the explanatory power of the framework.

\subsection{Synopsis}

Table~\ref{tab:cases} summarizes the three cases by the categories of the analysis scheme. The first two substantiate the two break directions of the uniqueness condition at different levels. The third stands deliberately outside the class.

\begin{table}[htbp]
\centering
\small
\begin{tabularx}{\textwidth}{L{2.2cm} L{4.2cm} L{1.6cm} L{3.2cm} L{2.6cm}}
\toprule
\textbf{Case} & \textbf{Break direction} & \textbf{Level} & \textbf{Effect of canonicalization} & \textbf{Relation to Section 2} \\
\midrule
ECDSA malleability & S-break / object-side multiple representation (multiple codes for one object) & Value & Failure class structurally excludable & $\mathrm{dec}^{-1}(\{x\})$ not a singleton \\
\addlinespace
Nomad & I-break / code-side semantic collapse (one code for multiple objects) & State/ storage & Failure class eliminable, tests separately needed & missing sentinel separation \\
\addlinespace
Wormhole (contrast) & none (missing account validation) & --- & Canonicalization does not apply & uniqueness condition not touched \\
\bottomrule
\end{tabularx}
\caption{The three case studies by the categories of the analysis scheme. The first two substantiate the two break directions at different levels. The third stands deliberately outside the class.}
\label{tab:cases}
\end{table}

The table makes the organizing principle visible: two structurally different failures, occurring in different systems and at different technical levels, prove to be the two directions of \emph{one} violated condition, while a third, superficially similar case falls cleanly outside. Precisely this sharpness of separation---the same root for the one, a clear boundary to the other---is what turns a collection of incidents into a class. The following sections first deepen how the canonicalization-guided version of the two instances concretely differs from the real specification (Section~5), and then place the class in the larger, cross-ecosystem context (Section~6).

\section{Specification and Canonicalization-Guided Version}

The case studies have shown \emph{that} the two instances violate the uniqueness condition. This section shows how narrow the difference is between the real specification and a canonicalization-guided version, and that in both technically very different cases the same abstract requirement appears as just one small change. We therefore contrast each case with its real specification, the break point, and the corrected version, and in the text name only the decisive condition in each case. The execution-near versions and their respective artifact status are given in the appendix: the Python version is tested, the Solidity version marked as compiled but not execution-tested.

\subsection{ECDSA: from ``valid'' to ``exactly one valid code''}

The real encoding rule that applies in Bitcoin with strict DER encoding restricts the \emph{form} of the two signature numbers, not their \emph{values}. Expressed as an acceptance condition, it requires well-formed encoding and values in the valid range, but admits both $s$ and the mirrored $n - s$. Precisely herein lies the break point: two values, the same signature, both accepted.

The canonicalization-guided version adds a single condition, the restriction to the lower half:
\[ s \leq n/2 \]
More is not needed. The additional line turns the acceptance rule into a canonical one: for each signature exactly one of the two mirrored forms remains valid. Decisive here is that the non-canonical form is \emph{rejected} and not silently recomputed. A recomputation---automatically mapping the upper value onto the lower---would be deterministic, but it would conceal that two forms existed, and would let the second persist via the detour of acceptance. It is the rejection that actually establishes the uniqueness condition. This distinction between deterministic recomputation and canonical rejection is precisely the separation of \emph{deterministic} and \emph{canonical} made in the foundational paper [1, Section 3.4]. This rejection logic concerns the verification of already bound, external signature bytes: there, a second form must not be accepted after the fact. In the \emph{generation} of a signature input, by contrast, an explicit normalization before the cryptographic binding is not only admissible but often the right canonicalization: thus, for instance, deterministic CBOR deliberately brings semantically equal numeric values to identical bytes before hashing. Rejection and normalization are therefore not in contradiction. They act at different points of the signature flow.

What additionally remained: the condition $s \leq n/2$ removes the malleability stemming from the mirrored $s$. It is a representation decision and not a statement about the security of the scheme. The further measures named in Section~4.1 remain unaffected by it.

\subsection{Nomad: from ``default value as state'' to ``sentinel outside the value range''}

The real checking logic decided on the validity of a root by reading out a confirmation time and returning ``valid'' as soon as this was non-zero and reached. The break point lies in the fact that the absence case (an unset entry) produces the same code \texttt{bytes32(0)} as the value confirmed at initialization, and that the storage model returns for the absence case not an error but precisely this zero value. Absence and validity collapse onto one code.

The canonicalization-guided version separates the two by excluding the absence code from the range of valid states. The supporting condition reads, put briefly:
\begin{quote}
the zero value \texttt{NULL\_ROOT} is never a confirmed state: by invariant, not by default value
\end{quote}
Concretely, one replaces the check that relies on a non-zero timestamp with an explicit confirmation marker that is structurally excluded for the zero value. An unknown entry then yields the default value ``not confirmed'' instead of a valid state. With this, the collapse between ``unknown'' and ``confirmed'' can no longer arise. Here too the correction is small and lies entirely at the representation level: it concerns the meaning of the zero value, not the logic of confirmation itself.

What additionally remained: as set out in Section~4.2, sentinel separation excludes the failure class, but replaces neither invariant tests nor taking the relevant audit findings seriously.

\subsection{One Principle, Two Small Changes}

The two versions come from technically far-apart worlds: the one from the number theory of elliptic curves, the other from the storage semantics of a virtual machine. Nevertheless, each adds exactly the same abstract requirement: that to each semantic object belongs exactly one valid code, in one case as an upper bound for $s$, in the other as exclusion of the zero value from the range of valid states. That one and the same principle appears in two such different systems as a single condition each is an indication that the uniqueness condition not only describes after the fact, but as a design heuristic suggests concrete checking conditions: in one case the upper bound for $s$, in the other the exclusion of the zero value from the range of valid states. The following section shows that this principle recurs far beyond the two worked-out cases, and places it in the larger context of already known precursors.

\section{The Grid and the Cross-Ecosystem Unification}

The preceding sections developed the vulnerability class on two worked-out cases. This section shows that the same class reaches far beyond these cases---and at the same time places it in the context of already known precursors. The paper's own contribution lies, to anticipate, not in the discovery of the phenomenon, but in its cross-ecosystem consolidation and in its connection with a computability-theoretic foundation.

\subsection{The Grid}

Combining the two break directions from Section~2 with the level at which uniqueness breaks yields a grid. As levels we distinguish the individual value, the composition of structured data, and the state or storage model. Table~\ref{tab:grid} enters documented cases into this grid.

\begin{table}[htbp]
\centering
\small
\begin{tabularx}{\textwidth}{L{2.6cm} L{5.3cm} L{5.3cm}}
\toprule
\textbf{Level $\downarrow$ / Direction $\rightarrow$} & \textbf{Object-side multiple representation (S-break)} & \textbf{Code-side semantic collapse (I-break)} \\
\midrule
Value & ECDSA malleability ($s$ / $n{-}s$; with \texttt{ecrecover} additionally via $v, r, s$) & \texttt{ecrecover} returns the zero value \texttt{address(0)} for an invalid signature, coinciding with a default or special value (Polygon MRC20, see below) \\
\addlinespace
Structure / serialization & multiple valid field-value encodings of the same semantic value (EIP-712, see below) & collision of concatenated dynamic fields without separation (non-injective packed encoding) \\
\addlinespace
State / storage & (open --- no documented case that would be a clean surjectivity break at this level) & default value = absence and validity (Nomad) \\
\bottomrule
\end{tabularx}
\caption{Grid of break direction and the level at which uniqueness breaks, with documented cases. The grid is an ordering device, not a strict partition; individual cases can be assigned to several cells.}
\label{tab:grid}
\end{table}

The cells are not equally densely populated, and some cases can, with good reason, be assigned to several cells. The grid is an ordering device, not a strict partition. The EVM primitive \texttt{ecrecover} is an instructive example of this: it shows both break directions at once: the signature malleability as object-side multiple representation and the fallback to \texttt{address(0)} for an invalid signature as code-side semantic collapse, unless the zero value is caught separately. Meant here is the unsafe raw use of the primitive. Audited libraries reduce this risk by explicitly validating signature form and recovery result (such as the restriction to low-S), yet the contract logic must nevertheless treat the zero value \texttt{address(0)} as an inadmissible sentinel where it would be security-relevant. The Polygon MRC20 incident (December 2021) substantiates this break direction with a real case: since \texttt{ecrecover} returns \texttt{address(0)} for an invalid signature and this return value was not checked, an execution in the name of the zero value could be brought about without a valid signature. The one code \texttt{address(0)} here carried at once the meaning ``signature error'' and that of a regular \texttt{address} value in the return type. This became dangerous because the return value was passed on unchecked as the sender address [21]. The case carries the observation made for the state level in the Nomad case onto the value level, but at the same time underscores the multi-causality of this framework: the full loss---around 800{,}000 stolen and around 9.2 billion endangered MATIC---became possible only through a second, independent gap, a missing balance check in the transfer path, which the operators themselves name as the core cause. Sentinel separation (rejection of \texttt{address(0)}) would have excluded the code-side semantic collapse, but not replaced the missing balance check. The purpose of the grid is to make visible that technically very different failures have a few common forms, and that the two worked-out cases (ECDSA malleability as multiple representation at the value level, Nomad as semantic collapse at the state level) mark the extreme points of a continuous structure. Two kinds of gap are to be kept apart here: the open cell \emph{within} this grid concerns object-side multiple representation at the state and storage level, for which no clean case is known to us. To be distinguished from this is the evidence situation \emph{outside} the grid, that is, the question of which of these cells are already tracked under their own name as a failure class in other ecosystems. There, as Section~6.2 shows, an asymmetry to the disadvantage of code-side semantic collapse is present, although its cell in the grid is occupied by Nomad. Appendix~C provides for central occupied cells a minimal, verifiable example that shows the underlying ambiguity at the byte or number level.

\subsection{The Same Pattern, Many Names}

The actual finding becomes visible when one looks beyond a single ecosystem. The same representation divergence has been described in a series of mutually independent fields, in each case under its own, locally coined name and mostly without systematic reference to the others. Table~\ref{tab:ecosystems} assembles these findings.

\begin{table}[htbp]
\centering
\small
\begin{tabularx}{\textwidth}{L{3.0cm} L{6.5cm} L{3.5cm}}
\toprule
\textbf{Field / ecosystem} & \textbf{Local name / reference case} & \textbf{Break direction} \\
\midrule
Bitcoin & Transaction malleability & S (signature/encoding) \\
Cosmos SDK & Non-deterministic value encoding & S (field value) \\
Ethereum / EIP-712 & (no class name of its own in the EVM context) & S (field value) \\
ABI / calldata & non-strict ABI decoding [7] & S (serialization) \\
Chain-agnostic (CAIP-380) & ``message malleability'' (threat model of Portable Proof) & S (message) \\
Software supply chain / Git & Hash chain malleability & S (signed commit representation: signature, OpenPGP subpacket, DER length) \\
Zero-knowledge systems & Non-canonical field-element representations (``aliasing'') [23]; related non-uniqueness of underdetermined circuits [24] & S (field value / public input; further cases context-dependent) \\
Ethereum / EVM (\texttt{ecrecover}) & Zero-value fallback \texttt{address(0)} (Polygon MRC20) & I (error and special value collapse) \\
\bottomrule
\end{tabularx}
\caption{Cross-ecosystem findings of the same vulnerability class, ordered by local name and break direction (S = object-side multiple representation, I = code-side semantic collapse). The density of evidence is asymmetric between the directions (Section 6.2).}
\label{tab:ecosystems}
\end{table}

Two entries deserve a more precise classification, because they show the breadth and at the same time the limits of the consolidation. The Cosmos case is especially instructive: the architecture document ADR-076 describes, under the name ``non-deterministic value encoding'', explicitly that semantically equal values, for instance with leading zeros or a differing decimal representation, can receive different encodings, and links this to faulty replay protection when the state mechanism relies on transaction hashes [8]. This is, in other terminology, exactly the object-side multiple representation at the field-value level. The Git case in turn shows that the pattern is not restricted to blockchains: a work on the software supply chain coins the term ``hash chain malleability'' and cites as its cause explicitly the inherent malleability in the data representations of a commit. One of its malleation routes is the ECDSA symmetry known from Section~4.1, another the non-canonical re-encoding of DER length fields [9]. That the same root was independently rediscovered in an entirely different field is itself the strongest evidence for the consolidation advocated here.

Two more recent entries fit the same pattern. The ABI case concerns the serialization level: Solidity's ABI encoder always produces the strict form, but the decoder does not enforce this strict mode, so that the same semantic calldata possesses several valid encodings, because a dynamic field may begin at a non-minimal offset. An authorization check that relies on a fixed calldata position can thus be circumvented with a non-canonical but valid encoding [7]. This is an object-side multiple-representation break at the composition level. The chain-agnostic case in turn shows that the pattern is already being addressed preventively: the threat model of the draft CAIP-380 (``Portable Proof'') lists ``message malleability'' as a point of its own and demands as a countermeasure explicitly a deterministic signer string and canonical JSON bytes before binding [10]: precisely the canonicalization before the cryptographic binding that Section~7 formulates as an obligation.

Not every case neighboring in the literature belongs cleanly in the same class, however. The analysis of the agentic payment protocol x402, for instance, names five attacks across several layers, among them missing replay protection [11]. This touches representation questions at the boundary between HTTP authorization and chain settlement, but is primarily a cross-layer security analysis and not a pure instance of the class considered here. We therefore present it as a borderline case, not as evidence, in the same stance with which Section~4.3 excludes the Wormhole case.

One property of this collection of evidence deserves explicit disclosure, because it concerns the reach of the framework: the evidence situation is markedly asymmetric between the two break directions. The vast majority of the findings listed in Table~\ref{tab:ecosystems} are object-side multiple representations. The nearby precursors from the classical security literature (Section~6.3) likewise concern predominantly this direction. For the code-side semantic collapse there is indeed, with the Polygon MRC20 incident, a documented case with real loss, and the Nomad case provides a worked-out I-break at the state level. Both, however, are our classification of concrete incidents, not a reference class already tracked under its own name, as the S-direction possesses many times over with terms such as ``transaction malleability'' or ``non-deterministic value encoding''. The S-direction is thus densely substantiated both by incidents and by established ecosystem terms, while the I-direction is substantiated by individual cases but conceptually hardly systematized.

This asymmetry does not refute the framework, but it limits what the collection of evidence carries. At the same time it can be read productively: if the uniqueness condition can break in only two ways, but practice has named almost exclusively the one direction, then the other marks a search zone. The code-side semantic collapse (default values that coincide with valid values, error returns that are read as valid objects, absence that is indistinguishable from validity) is accordingly to be treated at least as a less well systematized search zone. Whether it also occurs more rarely than the opposite direction is an empirical question that we do not answer here. The grid thus does not claim here to have an equally dense evidence situation. It predicts where one would be expected.

\subsection{The Precursors}

The principle behind all this is not new, and it would be dishonest to present it as such. The weakness catalogs addressed in the introduction list the subvertibility of checks through multiple representations of the same object under several active entries, such as the requirement not to validate before canonicalization, the faulty resolution of equivalent paths, and authorization based on non-canonical addresses [2]. Especially pertinent is the older line of signed structured data: the XML signature wrapping attacks allow a signed document to be restructured such that the signature check reads a different subtree than the processing application logic, without the signature becoming invalid [12]. An analysis of fourteen SAML frameworks found eleven vulnerable [13]. The cause description there---a gap between the object checked by the signature and the object processed by the business logic---is, in other language, exactly the canonicalization boundary named in Section~7. The case is thus a classical precursor of object-side multiple representation at the structure level. Likewise, the necessity of deterministic and canonical serialization is broadly developed in format standards: from strict DER encoding through the deterministic profiles of CBOR to canonical JSON schemas and the more recent, formally verified parsers for ASN.1 and CBOR [3, 4, 14]. One of these profiles, deterministic CBOR, even addresses the field-value level explicitly: it reduces semantically equal values---several representations of the same number, different forms of the zero value---to a canonical byte sequence and requires a Unicode normal form for text strings [15]. That deterministic serialization does not thereby already mean canonicity is stated explicitly by the Protobuf documentation, that deterministic output is not canonical, and the Cosmos sister document ADR-027 makes this precise as a missing bijectivity of the Protobuf encoding [16, 17]. Both confirm the separation of deterministic and canonical made in Section~2.1, on one of the most widely used serialization formats.

These precursors each anticipate part of the picture, and some do so with considerable formal depth: the deterministic serialization profiles and the verified parsers are not mere format technique, but address non-malleability partly with formal guarantees and, in the case of deterministic CBOR, explicitly at the field-value level. What they lack for the contribution pursued here is therefore not formality as such, but two specific things. First, they remain within their respective frame: the weakness catalogs treat canonicalization as a parsing and checking problem, the format and parser works as a question of format and schema design. None connects the individual cases across ecosystems into a class with a common organizing principle that engages with the hash, signature, and consensus identity of cryptographic systems. Second, the placement of this security phenomenon in the representation- and computability-theoretic frame is missing: that canonicalization is in the general case undecidable, that semantic equality of representations meets a computability boundary, is the subject of the representation theory of the foundational paper [1], yet in the security literature on these incidents this boundary is not drawn. The verified parsers, for instance, decide the well-formedness of a format, not the application-defined semantic equality of two field values of a Solidity or EIP-712 type, as it equates ``USDC'' and ``usdc''. The difference can be stated precisely: deterministic CBOR solves the problem for a concrete format profile, and the verified parsers check the schema well-formedness for CBOR and CDDL. The review procedure developed in Section~7 engages a level above: it asks whether the downstream application equality of a cryptographically bound field is represented at all by the bound bytes. This question precedes the format and is independent of it. It is the point at which the present work engages.

That the uniqueness condition not only orders failures after the fact, but is realizable as a design principle, is shown by a positive control case from the same domain as the case studies. The Binary Canonical Serialization (BCS) that arose in the Diem context, today used in several Move-based systems, makes canonicity a core property: it requires for each value of a type exactly one valid representation and rejects non-canonical encodings---such as non-minimal length prefixes---at deserialization, instead of silently recomputing them [18]. Its specification thereby formulates almost verbatim two properties that this work highlights as central: the uniqueness condition and byte-input invariance, according to which the signature of a message may be defined equivalently as the signature of the serialized bytes or of the in-memory value. BCS is thus not a counterexample but a confirmation: where the uniqueness condition is deliberately institutionalized, the failure class examined here does not arise at the serialization level in the first place. What BCS accomplishes at the format and composition level, the review procedure of the following section lifts onto the application-defined field-value level, at which a format alone can no longer fix equality.

\subsection{The Contribution as a Bridge}

With this, the contribution of this section can be stated precisely: the connection of two hitherto separate perspectives. On the one side stands classical canonicalization security, which knows the attack pattern but treats it as a local parsing problem. On the other side stands representation and computability theory, which provides the foundation (the notion of the uniqueness condition, the distinction of deterministic and canonical, byte-input invariance), but was developed without reference to concrete attacks. Between the two runs the bridge that this work builds: the scattered, named incidents can be read as instances of \emph{one} violation of the uniqueness condition, in one of two directions. Their recurring occurrence is explained by the fact that this representation-theoretic root has so far not been tracked as a common, cross-ecosystem grid in the blockchain and crypto-protocol security literature, not by the principle of canonicalization having been unknown anywhere.

Within this consolidation lies a smaller, concrete observation, which we name as an instance of its own without overstating it. In the Ethereum context of structured, signed data per EIP-712, the \emph{structure} of the encoding is carefully fixed, but the \emph{canonicity of the field values within} this structure is not throughout. If a signed data object contains a field whose type admits several semantically equivalent but byte-different values (such as a string interpreted on the application side without regard to upper and lower case), then each of these forms produces a different structure hash. Unlike the ECDSA malleability from Section~4.1, where a second signature can be derived from one without knowledge of the key, here no second code arises through signature transformation: each variant produces a different signature input, and if the signer or a signature workflow admits several semantically equal variants, then several valid signatures arise for the same intent. The failure thus lies at the message and field-value level, before hashing, not in the signature itself. A protection that tracks already processed structure hashes or byte-wise messages instead of a canonical semantic object identifier does not apply, because each variant appears as a new message. A per-signer counter bound globally, before the semantics, can by contrast prevent the concrete reuse, whereby the field-value ambiguity then persists as a multi-signature and idempotency problem. This is an object-side multiple representation at the field-value level, and in the EVM context it is, to our knowledge, typically not tracked as a class of its own, separate from replay and signature malleability. Important is the precondition that limits the case: it applies only where an application-defined type admits value ambiguity. For an already canonical type such as a fixed-width integer, the problem does not arise. This instance is thus neither surprising nor new in principle: it is exactly what the framework predicts at this point of the grid, and a deterministic encoding profile such as the one named above would already exclude it at the format level. In the application-specific EIP-712 context, it nevertheless remains open.

The value of the consolidation thus lies less in individual entries of the grid than in the explanation it provides: why a failure class long known in principle reappears in every new cryptographic ecosystem and each time receives a name of its own, and how a representation-theoretic perspective, applied early, could reduce the exploitability of these systems.

\section{From the Framework to the Review Procedure}

The framework so far orders and explains, but it is not yet a tool. This section derives from it an applicable review procedure: not a new theoretical result, but a translation of the insight into checkable design questions. The benefit lies in the fact that a developer or reviewer can thereby decide \emph{before} the damage whether a system is exposed to the failure class described.

\subsection{Canonicity Is Relative to an Equality Relation}

The starting point is a clarification that forestalls an obvious objection. One might object that bytes are, after all, always unique: the same byte sequence is the same byte sequence. This is correct and at the same time misses the point. The security problem arises not at the byte level, but where an application uses a \emph{coarser} equality than the byte-wise one: where it treats two different byte sequences as ``the same''. Canonicity is therefore not an absolute notion, but always relative to a semantic equality relation. Without a fixed such relation, canonicity is not well-defined, and conversely the risk arises precisely when this relation is coarser than the byte-wise equality at which hash and signature engage.

\subsection{The Canonicalization Obligation}

From this observation follows a single, checkable rule, which we formulate as the \emph{canonicalization obligation}:
\begin{quote}
Every cryptographically bound representation must, for every downstream semantic equality relation, either enforce a canonical normal form or explicitly exclude this equality relation from the security context.
\end{quote}
More briefly: \emph{Sign what you compare, and compare what you sign---after canonicalization.} A break of this obligation is precisely the vulnerability class of this work: if binding occurs over bytes (hash, signature, replay key, consensus identity) but comparison occurs at a coarser semantic level, without the bytes having first been brought to the canonical form of this level, then representation divergence arises.

The obligation can be transferred into a short sequence of review steps that can be applied mechanically:
\begin{enumerate}
\item What is the cryptographically bound byte object---what exactly is hashed, signed, used as a replay key or consensus identity?
\item Which semantic equality does the application subsequently use on this object or its components?
\item Is this semantic equality identical to byte-wise equality? If yes, there is no canonicalization risk from this relation.
\item If no: does a \emph{computable, normatively specified, and implemented} canonical normal form exist for this equality? A merely mathematically existing normal form that is not specified or not implemented does not suffice: the protocol must actually be able to enforce it.
\item Is exactly this normal form enforced \emph{before} hashing or signing? If no, a canonicalization risk is present.
\end{enumerate}

The value of this sequence is that it translates the abstract framework into a series of concrete, answerable questions: it turns the taxonomy into a review procedure. What this looks like is shown by a run through the EIP-712 case from Section~6.4, for a field \texttt{asset} that carries a token symbol:
\begin{enumerate}
\item \emph{Bound byte object:} the structure hash over the signed object, into which the field value \texttt{asset} enters.
\item \emph{Downstream equality:} the application compares token symbols without regard to upper and lower case---``USDC'' and ``usdc'' denote the same object.
\item \emph{Byte-wise equality?} No: the two forms are byte-different but are semantically equated. The relation is coarser than the byte-wise one.
\item \emph{Computable, specified, implemented normal form?} Yes---such as a fixed case reduction or, for the full character set, a specified Unicode normal form.
\item \emph{Enforced before binding?} In the observed practice, no: the structure hash is formed over the unchecked field value. Thus a canonicalization risk is present, and the check ends with a finding.
\end{enumerate}

The sequence here yields, in five steps, exactly the diagnosis that Section~6.4 develops argumentatively, and at the same time names the point at which the correction must engage. The appendix shows the corresponding remediation.

Its status is to be delimited precisely: the canonicalization obligation is not a new theorem and not a complete decision procedure for arbitrary semantics, but a soundness-oriented review obligation. Whoever uses a coarser semantic equality than byte-wise equality must either specify a computable normal form and enforce it before the cryptographic binding, or keep the equality out of the security-relevant path. The obligation says \emph{what} is to be checked and \emph{when} a risk is present. It does not automatically decide whether a given semantics possesses a computable normal form: this question can, in the general case, meet the computability boundary described in the foundational paper.

For individual format profiles, the same obligation is already normatively expressed. The JSON Canonicalization Scheme justifies itself explicitly by the fact that cryptographic operations such as hashing and signing require an invariant data representation, and for this purpose fixes a canonical JSON form [3]. What is standardized there for a format, the canonicalization obligation generalizes to the application-defined semantic equality that precedes a format and is not fixed by it.

\subsection{Two Levels of Composite Objects}

For structured objects, canonicity splits into two conditions that can be violated independently of one another. \emph{Composition canonicity} asks whether the fields of a composite object are uniquely separable, that is, whether the byte sequence can be uniquely decomposed into its components. \emph{Value canonicity} asks whether each individual field value has exactly one representation. Relative to a fixed field structure and semantic equality relation, both are necessary: a structured cryptographic object is canonical only if both its composition is uniquely decomposable and each contained field value is canonical.

This decomposition classifies two otherwise seemingly unconnected cases: the collision of concatenated dynamic fields without separation violates composition canonicity, while the ambiguous field-value encoding---the EIP-712 case from Section~6.4---violates value canonicity. We emphasize that both levels are individually treated in the literature: unique decomposability as the classical question of unique decodability, field-value normalization in deterministic serialization profiles, which even explicitly name the interaction of both levels---for instance that a value reduction can affect the uniqueness of mapping keys [15]. The decomposition here therefore serves not as a new result, but as an ordering device that names precisely the gap between the composition level and the field-value level.

\subsection{A Field-Type Classification for Signed Data}

The canonicalization obligation can, for practice, be broken down to the field types of signed data. Table~\ref{tab:fieldtypes} classifies common Solidity and EIP-712 field types by their canonicity risk, not as a conclusive norm, but as a starting point for the sequence of review steps from 7.2.

\begin{table}[htbp]
\centering
\small
\begin{tabularx}{\textwidth}{L{2.8cm} L{8.0cm} L{2.2cm}}
\toprule
\textbf{Field type} & \textbf{Canonicity status} & \textbf{Risk} \\
\midrule
\texttt{uint256}, \texttt{bool} & intrinsically canonical if the semantics is byte- or value-equal & low \\
\addlinespace
\texttt{bytes32} & intrinsically canonical if really meant as an atomic 32-byte value; risky if used as a packed container of several fields & low / medium \\
\addlinespace
\texttt{address} & canonical as 20 bytes; risky with string representation, ENS/alias resolution, or case-insensitive UI/API processing & medium \\
\addlinespace
\texttt{string} & not canonical without a fixed normal form (such as Unicode normalization, upper/lower case) & high \\
\addlinespace
\texttt{bytes} & low if opaque and compared byte-wise; high as soon as later interpreted as a structured, normalized, or schema-bound object & low / high \\
\addlinespace
Arrays & canonical as a sequence; risky if the semantics is set or multiset (order irrelevant) & high \\
\addlinespace
Decimal or number strings & risky without a numeric normal form (leading zeros, decimal places, exponential form) & high \\
\bottomrule
\end{tabularx}
\caption{Canonicity risk of common Solidity/EIP-712 field types as a starting point for the sequence of review steps from Section 7.2.}
\label{tab:fieldtypes}
\end{table}

The table operationalizes the core statement: a field is uncritical exactly when its application-side equality coincides with the byte-wise one. As soon as the application normalizes a field, compares without regard to upper and lower case, sorts, interprets numerically, or treats it as a set, the byte-wise binding (signature, hash) recedes behind a coarser semantic equality, and the canonicalization obligation requires a normal form enforced before the binding. For the string case, the normal form needed for this is already normatively worked out: the Unicode normalization forms [19] fix the canonical composition, and the Unicode security mechanisms [20] additionally recommend case-folded forms to reduce confusion risks. Decisive is the order, which the platform documentation also emphasizes: the normalization must precede the security-relevant check and the cryptographic binding, not follow it: exactly the position marked in Figure~\ref{fig:boundary}, before the canonicalization boundary.

\subsection{The Canonicalization Boundary as an Operational Model}

The three preceding building blocks can be summarized in a single flow model that traces the path of a value from the raw input to the cryptographic binding and locates the possible break points along it.

The model makes the failure classes visible as positions relative to a \emph{canonicalization boundary}. The two break directions from Section~2 appear in it as positions in the flow: if hash or signature is formed \emph{before} the canonicalizer, the object-side multiple representation (S-break) arises, because several presentations of the same object yield different bound bytes. An ambiguous decoder produces the code-side semantic collapse (I-break), because different objects fall onto the same code. This break sits at the reading step itself and is therefore not to be healed by an upstream normalization: it is a defect of the decoding mapping, which only an injective redefinition of the assignment, such as the exclusion of the sentinel from the value range, remedies.

From these two, a third, subordinate phenomenon is to be distinguished, which we call an \emph{idempotency break}: if comparison occurs on the semantic object but a replay or deduplication key is kept on the bytes, semantically equal messages count as different and are processed multiple times. The idempotency break is not a third break direction of the uniqueness condition (the dichotomy from Section~2.2 remains complete), but a practical consequence of the same missing canonicalization at the key level: it arises because the deduplication key is formed before the semantic normalization. We list it here because in practice it is the most frequent effect of an S-break on replay protection mechanisms.

\begin{figure}[htbp]
\centering
\includegraphics[width=\textwidth]{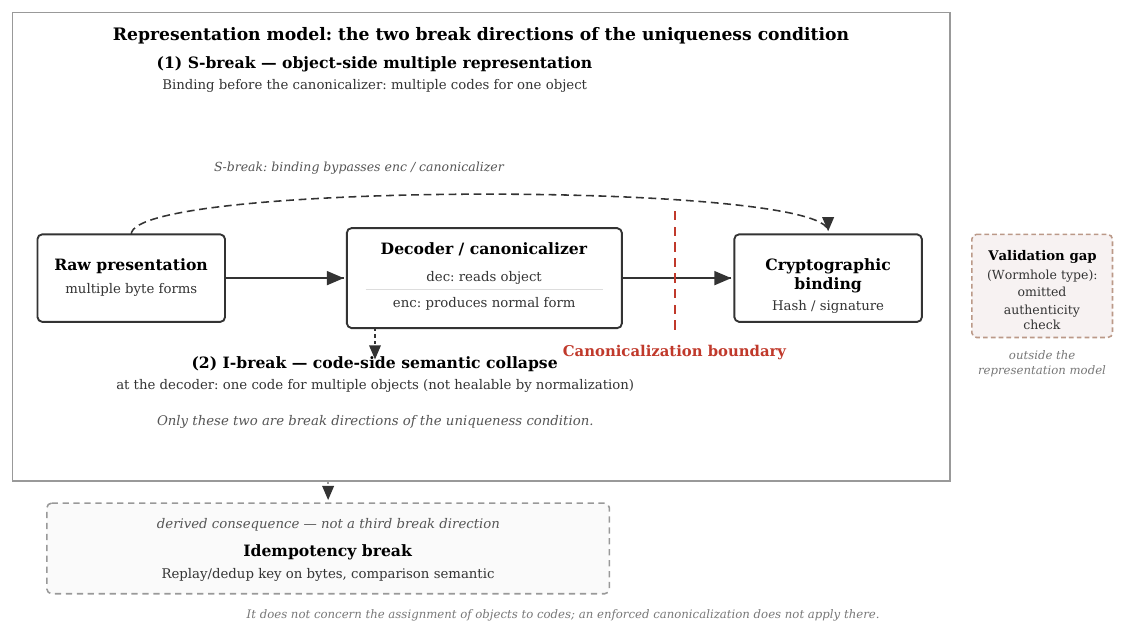}
\caption{Canonicalization boundary and break points of cryptographic binding. The two break directions (S-break, I-break) are positions relative to the canonicalization boundary; the idempotency break is a subordinate consequence at the key level, and the validation gap lies outside the representation model.}
\label{fig:boundary}
\end{figure}

The validation gaps treated in Section~8---an omitted authenticity check as in the Wormhole case---by contrast lie entirely outside this model, because they do not concern the assignment of objects to codes, but an entirely missing check. Figure~\ref{fig:boundary} summarizes this placement. The model is thereby at once a tool and a boundary marker: it shows where the framework applies, and where a failure lies outside its domain.

\section{Where Representation Divergence Is Not the Cause}

A framework gains credibility less through the cases it explains than through those it explicitly excludes. This section names two areas in which the representation-theoretic perspective does \emph{not} carry---not as an aside, but as a necessary delimitation of the claim.

The first area is the Wormhole incident already treated in Section~4.3, and with it the entire class of validation gaps. Where a check that should have taken place simply fails to occur (a missing signature control, an unverified account identity, an omitted authorization check), there is no representation problem. There is no second valid code for the same object there, and no collapse of two meanings onto one code. The uniqueness condition is not touched at all. Such cases are often similarly expensive and superficially similar to those treated here, and precisely for that reason the delimitation is important: an enforced canonicalization would not have prevented them, because their cause lies not in the assignment of objects to codes.

The second area concerns non-deterministic arithmetic, such as rounding differences between different processor architectures or floating-point implementations. At first glance this seems to be a representation question, the same computational result in different bit patterns, and in systems that rely on byte-wise agreement of computational results, a divergence can indeed arise from it. For the consensus-relevant arithmetic of the systems considered here, however, this vector is largely defused, and deliberately so: many large smart-contract and ledger environments, in particular the EVM context, avoid floating-point arithmetic for consensus-critical computations and use integer or fixed-specified arithmetic precisely because the non-determinism of floating point is known. Where floating point nevertheless occurs, such as in off-chain computations or in certain execution environments, the resulting divergence is more a problem of reproducibility than one of canonicity in the sense developed here. We therefore present the case in order to delimit it: it shows that not every ambiguity at the byte level is a violation of the uniqueness condition, and that the framework ends where the ambiguity stems not from the representation of an object, but from the non-reproducibility of a computation. The contrast with the foundational paper sharpens this boundary: there, an irrational value such as $\pi$ produces divergent bytes precisely when its presentation is not normatively fixed [1, Section 7.3], a representation problem that a canonical specification remedies. The floating-point divergence delimited here is its counterpart: it arises not from the missing fixing of a presentation, but from the non-reproducibility of the computation itself, and therefore remains unremediable by canonicalization.

A third, more general point belongs to this delimitation: not every missing normalization is a security problem. A non-canonical representation becomes security-relevant exactly when a security-critical decision hangs on it: that is, when a hash, a signature, a replay or deduplication key, an authorization decision, or a consensus identity is formed over the bytes, while the application compares the associated objects at a coarser semantic level. If this coupling is absent, the missing normalization remains a matter of data hygiene: two forms of the same value may then be inconsistent, but no check, no binding, and no state decision is subverted. The criterion is thus not the ambiguity itself, but its coupling to a byte-wise binding with coarser semantic evaluation. The framework of this work targets only this constellation.

Both delimitations serve the same purpose. The perspective proposed here explains a particular, clearly delineated class of failures and claims no more. Its usefulness depends on this boundary remaining visible, on a user of the framework being able to recognize when an incident falls into its domain and when it does not.

\section{Limits}

The preceding sections have already delimited the claim of this work at several points. We summarize the limits here, because their clear naming is, for a contribution of this kind, not embellishment but substance.

\textbf{Category of the contribution.} This work introduces no new cryptographic primitive and proves no new impossibility result. Its contribution is systematization, transfer, and the connection of two hitherto separate perspectives---classical canonicalization security and representation theory. Neither the security phenomenon nor the mathematical principle is, taken on its own, new. New is their consolidation as a cross-ecosystem grid for cryptographic and distributed systems, supplemented by the review procedure derived from it in Section~7. Measured by the standards of a systematization contribution, the work is therefore to be assessed by the quality of the ordering and its explanatory power, not by the novelty of individual building blocks.

\textbf{Empirical basis.} We illustrate the class with worked-out cases and a cross-ecosystem collection of evidence, but do not systematically survey how frequently the failure class occurs or how it distributes quantitatively across the two break directions and the levels. The collection of evidence shows that the pattern occurs in many fields. It is not proof of its relative significance compared with other failure classes. In addition, there is the asymmetry disclosed in Section~6.2: the object-side multiple representation is supported by numerous local ecosystem terms, while the code-side semantic collapse, though substantiated by documented incidents (Nomad, Polygon MRC20), is not tracked as a named reference class in the surveyed literature. A quantitative study over a corpus of documented incidents remains open and would be a worthwhile addition, in particular for the I-direction, which the grid identifies as a less well systematized search zone.

\textbf{Necessary, not sufficient.} Real security incidents are almost always multi-causal. Where we say that an enforced uniqueness condition would have reduced the exploitability of a failure class or excluded this class, it is never claimed that canonicalization alone would have prevented the entire loss. The Nomad case shows this exemplarily: sentinel separation excludes the failure class in question, but replaces neither invariant tests nor the appropriate response to audit findings. Canonicalization is a structural countermeasure against a particular class, not a comprehensive protection.

\textbf{No security promise.} Canonicity is a property of the representation, not of cryptographic security in the narrower sense. The perspective developed here makes no statement about secrecy, key security, or collision resistance of the primitives used. A system can be canonical in every respect and nevertheless be insecure for other reasons.

\textbf{Reactive and preventive.} It would be unfair to reproach the affected communities for having overlooked the problem. On the contrary: several of the incidents treated were responded to with exactly the right means: with the low-S normal form, with the separation of signature data from the transaction identity, with deterministic serialization profiles and verified parsers. The finding of this work is not that no one thinks about canonicalization, but that the canonical form is typically fixed normatively only \emph{reactively}, after an expensive incident, instead of being derived \emph{preventively} from the representation principle, and that a common, cross-ecosystem perspective together with the review procedure from Section~7 could help shorten this repetition. In this reading, the work is a contribution to prevention, not a criticism of the reaction.

\textbf{Dependence on a specified equality.} The review procedure from Section~7 works only insofar as the downstream semantic equality relation can be made explicit at all. Where an application does not clearly name its equality, for instance because it arises implicitly through scattered comparison logic, the canonicalization obligation cannot be applied mechanically, but first requires making this equality visible. That is a precondition, not a guarantee.

\textbf{False positives and design trade-offs.} Not every non-canonical input must be forbidden. The canonicalization obligation knows two admissible resolutions, namely to enforce a normal form \emph{or} to keep the equality in question out of the security-relevant path, and in some cases the second is the simpler: it can suffice to conduct security-critical comparisons exclusively byte-wise. A user of the framework must therefore distinguish between genuine risk and merely missing data hygiene. Too literal an application produces false alarms.

\textbf{Outlook beyond cryptographic bindings.} The perspective developed here presupposes a byte-wise binding on which a security decision hangs. The basic pattern---a surface form, an intended meaning, and a downstream instance that resolves the two differently---occurs, however, also outside cryptographic systems, for instance with inputs to language-processing models and the filters, tools, and agents downstream of them. Whether the canonicalization obligation carries there is an open question: the clearly specified semantic equality relation is missing, without which canonicity per Section~7.1 is not well-defined, and in place of a byte-wise binding comes a probabilistic interpretation. We therefore present the case explicitly as an outlook and not as an instance of the class substantiated here. A transfer would have to be developed and substantiated independently.

\section{Conclusion}

Cryptographic systems mean semantic objects, but they operate on bytes. Between the two lies a translation that is rarely unique, and where its uniqueness is not deliberately enforced, an attack surface opens up as soon as security decisions hang on the representation. We have shown that a series of expensive security failures, independently discovered and locally named in different ecosystems, has the same root: the violation of a single uniqueness condition, relative to a fixed equality relation, in one of two directions, that is, multiple codes for one object or one code for multiple objects.

The contribution lay in the consolidation of these scattered findings, in their connection with a computability-theoretic foundation, and in the review procedure derived from it. The classical security literature knows the canonicalization problem but treats it locally. Representation theory provides the foundation, but without reference to concrete attacks. The bridge between the two explains why the same failure class reappears in every new cryptographic ecosystem and each time receives a name of its own, and the canonicalization obligation from Section~7 translates this insight into a handful of checkable design questions, with which the risk can be recognized preventively instead of being fixed reactively after the damage.

With this, the arc back to the foundational paper closes. There it was shown that for the operational use of a value in a protocol, it does not suffice which value it has, but that decisive is through which normatively fixed presentation it becomes a reproducible operational object [1]. The present work is the security-side reverse of this insight: where this normative fixing is missing or incomplete, not only ambiguity arises, but a concrete, recurring vulnerability class. The guide in the appendix translates this insight into immediately applicable checking steps: from the foundation, through the insight, to the tool.

\appendix

\section{Developer Guide}

This appendix translates the framework into immediately applicable artifacts: a review checklist (the sequence of review steps from Section~7.2 in applicable form), a generic canonicalization template, and the canonicalization-guided versions of the two case studies. For reproducibility, we record the execution status of each code part explicitly: the Python artifacts are executable and are accompanied by a test that substantiates the central malleability property (four check cases, green); the Solidity artifacts are compilable against \texttt{solc} 0.8.26 but were not executed on a test chain for lack of an available EVM execution environment---they are therefore marked as \emph{compiled, not execution-tested}. Where this appendix and Appendix~B provide the applicable and executable material, Appendix~C adds a purely didactic counterpart that shows the failure class through minimal, verifiable examples.

\subsection{The Canonicalization Checklist}

The following list is the operationalized sequence of review steps from Section~7.2. It is run through once per cryptographically bound object:
\begin{enumerate}
\item \textbf{Identify the binding.} What exactly is hashed, signed, used as a replay key or as a consensus identity? Note the exact byte sequence that is bound.
\item \textbf{Name the downstream equality.} Which semantic equality relation does the application subsequently use on this object or its fields? Is anything normalized anywhere, compared without regard to upper/lower case, sorted, interpreted numerically, or treated as a set?
\item \textbf{Byte-wise equality?} Does this equality coincide with byte-wise equality? If yes, there is no canonicalization risk from this relation, step done.
\item \textbf{Normal form present?} If no: does a \emph{computable, normatively specified, and implemented} canonical normal form exist for this equality? A merely mathematically existing normal form does not suffice.
\item \textbf{Enforced before binding?} Is the normal form enforced before hashing or signing? Here the two paths are to be distinguished (Section~5.1): in the generation path, an object may be normalized before binding. In the verification path, already bound, externally presented bytes in non-canonical form must be \emph{rejected} instead of silently recomputed. If this is omitted, a canonicalization risk is present.
\end{enumerate}

Two admissible resolutions (Section~9): either enforce a normal form \emph{or} keep the equality in question out of the security-relevant path (such as conducting security-critical comparisons exclusively byte-wise). For composite objects, the check is to be conducted in two stages (Section~7.3): once for composition canonicity (unique decomposability), once for the value canonicity of each field.

\subsection{Generic Canonicalization Template (Python)}

The template separates the two points in the flow (Figure~\ref{fig:boundary}): in the \emph{generation} of a signature input, normalization is admissible (\texttt{normalize}). In the \emph{verification} of already bound, external bytes, a non-canonical form is \emph{rejected} (\texttt{accept}), not recomputed. The uniqueness condition is operationally enforced by having \texttt{accept} return an object exactly when the bytes are identical to \texttt{encode(decode(code))}.

\begin{lstlisting}[language=Python]
from abc import ABC, abstractmethod
from typing import Generic, TypeVar
T = TypeVar("T")
class NonCanonicalError(ValueError):
    """Bytes are validly decodable, but not the canonical form."""
class Canonical(ABC, Generic[T]):
    @abstractmethod
    def decode(self, code: bytes) -> T: ...
    @abstractmethod
    def encode(self, obj: T) -> bytes: ...

    def is_canonical(self, code: bytes) -> bool:
        try:
            obj = self.decode(code)
        except Exception:
            return False
        return self.encode(obj) == code
    def accept(self, code: bytes) -> T:
        """Verification path: accepts only the canonical form,
        rejects any second form (no silent recomputation)."""
        obj = self.decode(code)
        if self.encode(obj) != code:
            raise NonCanonicalError("non-canonical form rejected")
        return obj
    def normalize(self, obj: T) -> bytes:
        """Generation path: brings an object to its form before binding."""
        return self.encode(obj)
\end{lstlisting}

The rejection in \texttt{accept} is the supporting decision: a silent recomputation would be deterministic, but would let the second form persist via the detour of acceptance (Section~5.1).

\subsection{ECDSA: Low-S Canonicalization with Substantiating Test (Python)}

The following excerpt substantiates Section~4.1: for a valid signature \texttt{(r, s)}, \texttt{(r, n - s)} is also valid, and the low-S rule \texttt{s <= n/2} selects exactly one of the two forms as canonical. Generation normalizes; verification rejects high-S.

\begin{lstlisting}[language=Python]
from ecdsa import SECP256k1
from ecdsa.util import sigencode_string, sigdecode_string
N = SECP256k1.order
HALF_N = N // 2
BASELEN = SECP256k1.baselen  # 32
def is_low_s(s: int) -> bool:
    return s <= HALF_N
def split_sig(sig: bytes) -> tuple[int, int]:
    return (int.from_bytes(sig[:BASELEN], "big"),
            int.from_bytes(sig[BASELEN:], "big"))
def join_sig(r: int, s: int) -> bytes:
    return r.to_bytes(BASELEN, "big") + s.to_bytes(BASELEN, "big")
def canonical_low_s(sig: bytes) -> bytes:      # generation path: normalize
    r, s = split_sig(sig)
    return join_sig(r, s if is_low_s(s) else N - s)
def accept_canonical(sig: bytes) -> bytes:     # verification path: reject
    r, s = split_sig(sig)
    if not is_low_s(s):
        raise ValueError("non-canonical high-S signature rejected")
    return sig
\end{lstlisting}

The accompanying test checks four properties and runs green: (1) \texttt{(r, s)} and \texttt{(r, n - s)} both verify; (2) exactly one of the two forms is low-S; (3) \texttt{canonical\_low\_s} maps both onto the same byte sequence and is idempotent; (4) the verification path accepts low-S and rejects high-S. The complete test with all four check cases is in Appendix~B.1.

\subsection{Nomad: Sentinel Separation (Solidity)}

The contract implements Section~5.2: the zero value \texttt{NULL\_ROOT} never denotes a confirmed state, by invariant. Validity no longer hangs on a non-zero timestamp, but on an explicit \texttt{confirmed} marker whose default value for an unset entry is \texttt{false}. The code is compilable without error against \texttt{solc} 0.8.26; it was not executed on a test chain (\emph{compiled, not execution-tested}).

\begin{lstlisting}[language=Java]
// SPDX-License-Identifier: MIT
pragma solidity 0.8.26;
contract CanonicalReplica {
    bytes32 internal constant NULL_ROOT = bytes32(0);
    struct Confirmation {
        bool confirmed;      // supporting invariant, not the timestamp
        uint64 confirmedAt;  // purely informational
    }
    mapping(bytes32 => Confirmation) private _confirmations;
    error NullRootNotConfirmable();
    error AlreadyConfirmed(bytes32 root);
    function confirmRoot(bytes32 root) external {
        if (root == NULL_ROOT) revert NullRootNotConfirmable();
        if (_confirmations[root].confirmed) revert AlreadyConfirmed(root);
        _confirmations[root] = Confirmation(true, uint64(block.timestamp));
    }
    function isConfirmed(bytes32 root) public view returns (bool) {
        if (root == NULL_ROOT) return false;  // sentinel never valid
        return _confirmations[root].confirmed;
    }
    function isMessageProven(bytes32 messageRoot) external view returns (bool) {
        return isConfirmed(messageRoot);
    }
}
\end{lstlisting}

The decisive difference from the real version is small and lies entirely at the representation level: not a computational step is corrected, but the meaning of the zero value. Appendix~B.2 substantiates both executably.

\subsection{EIP-712: Field-Value Canonicalization (Solidity)}

The core implements Section~6.4: a field interpreted case-insensitively on the application side is brought to a canonical form \emph{before} the structure hash, so that ``USDC'' and ``usdc'' produce the same digest. A non-ASCII byte is rejected instead of silently passed through (rejection instead of recomputation, Section~5.1). The complete contract moreover checks the signature itself for its canonical form: it enforces low-S, restricts the recovery identifier, and catches the \texttt{address(0)} fallback of \texttt{ecrecover}. The complete contract is in Appendix~B.3 and is compilable against \texttt{solc} 0.8.26 (\emph{compiled, not execution-tested}).

\begin{lstlisting}[language=Java]
// keccak256("Order(string asset,uint256 amount,uint256 nonce)")
bytes32 private constant ORDER_TYPEHASH =
    keccak256("Order(string asset,uint256 amount,uint256 nonce)");
error NonAsciiByte();
// Canonical normal form of the case-insensitive field: ASCII lowercase.
function canonicalizeAsset(string memory asset) public pure returns (string memory) {
    bytes memory b = bytes(asset);
    for (uint256 i = 0; i < b.length; i++) {
        uint8 c = uint8(b[i]);
        if (c > 0x7F) revert NonAsciiByte();
        if (c >= 0x41 && c <= 0x5A) b[i] = bytes1(c + 32); // 'A'..'Z' -> 'a'..'z'
    }
    return string(b);
}
// Structure hash over the CANONICALIZED field value.
function hashOrder(string memory asset, uint256 amount, uint256 nonce)
    public pure returns (bytes32)
{
    string memory canonAsset = canonicalizeAsset(asset);
    return keccak256(abi.encode(
        ORDER_TYPEHASH,
        keccak256(bytes(canonAsset)),  // canonical field-value binding
        amount,
        nonce
    ));
}
\end{lstlisting}

Unlike the ECDSA malleability, the fix here lies before hashing, at the field-value level: it removes the ambiguity before it enters the structure hash.

\subsection{Integration Note}

The canonicalization belongs at exactly one place: immediately before the cryptographic binding, on the generation side, and as rejection on the verification side. If it is distributed across scattered comparison logic, the precondition named in Section~9 arises again, that of first having to make the downstream equality visible. A single, named canonicalization point per bound object keeps the sequence of review steps mechanically applicable. The complete versions of all three artifacts are found in Appendix~B.

\section{Complete Artifacts}

This appendix prints the complete artifacts shown only in excerpt in Appendix~A, so that the paper remains independently checkable. The two Python tests are executable and run green (four check cases each). They substantiate the two break directions. The Solidity contract is compilable without error against \texttt{solc} 0.8.26, but was not executed on an EVM test chain (\emph{compiled, not execution-tested}).

\subsection{Complete ECDSA Substantiating Test (Python)}

Execution: \texttt{pip install ecdsa pytest}, then \texttt{python3 -m pytest -q test\_ecdsa\_low\_s.py}. The test imports the functions shown in A.3 from \texttt{ecdsa\_low\_s.py}.

\begin{lstlisting}[language=Python]
import hashlib
from ecdsa import SigningKey, SECP256k1
from ecdsa.util import sigencode_string, sigdecode_string
from ecdsa_low_s import (
    N, is_low_s, flip_s, split_sig, join_sig,
    canonical_low_s, accept_canonical,
)
MSG = b"canonicalization is a representation property"
def _make_sig():
    sk = SigningKey.generate(curve=SECP256k1, hashfunc=hashlib.sha256)
    vk = sk.get_verifying_key()
    sig = sk.sign(MSG, hashfunc=hashlib.sha256, sigencode=sigencode_string)
    return sk, vk, sig
def test_both_forms_valid():
    """(r, s) and (r, n - s) are BOTH cryptographically valid."""
    _, vk, sig = _make_sig()
    r, s = split_sig(sig)
    flipped = join_sig(r, flip_s(s))
    assert vk.verify(sig, MSG, hashfunc=hashlib.sha256, sigdecode=sigdecode_string)
    assert vk.verify(flipped, MSG, hashfunc=hashlib.sha256, sigdecode=sigdecode_string)
    assert sig != flipped   # two different codes, the same signature
def test_exactly_one_low_s():
    """Of (r, s) and (r, n - s), exactly one is the low-S form."""
    _, _, sig = _make_sig()
    _, s = split_sig(sig)
    s2 = flip_s(s)
    assert is_low_s(s) != is_low_s(s2)   # exactly one, never both/neither
    assert (s + s2) == N                  # they are mirrored at n
def test_normalize_idempotent():
    """normalize brings both mirrored forms to the same low-S byte sequence."""
    _, _, sig = _make_sig()
    r, s = split_sig(sig)
    flipped = join_sig(r, flip_s(s))
    c1 = canonical_low_s(sig)
    c2 = canonical_low_s(flipped)
    assert c1 == c2                       # one object -> one canonical code
    assert is_low_s(split_sig(c1)[1])     # and it is low-S
    assert canonical_low_s(c1) == c1      # idempotent
def test_accept_rejects_high_s():
    """Verification path accepts low-S and REJECTS high-S (no recomputation)."""
    _, _, sig = _make_sig()
    low = canonical_low_s(sig)
    high = join_sig(split_sig(low)[0], flip_s(split_sig(low)[1]))
    assert accept_canonical(low) == low   # low-S accepted
    try:
        accept_canonical(high)
    except ValueError:
        pass
    else:
        raise AssertionError("high-S should have been rejected")
\end{lstlisting}

\subsection{Complete Nomad Sentinel Substantiating Test (Python)}

This test substantiates the code-side semantic collapse (Section~4.2) executably and shows that sentinel separation (Section~5.2) excludes it without breaking the legitimate function. It runs green (four check cases). \emph{Scope:} the test is not an EVM execution. It reproduces in Python the property of the storage model decisive for the case---an unset mapping entry returns the zero value instead of an error.

\begin{lstlisting}[language=Python]
NULL_ROOT = b"\x00" * 32          # bytes32(0) - absence/default value
UNKNOWN   = b"\xde\xad" * 16      # a never-registered message
class EVMMapping(dict):
    """The decisive EVM property: an unset entry
    returns the default value, not an error."""
    def __init__(self, default):
        super().__init__(); self._default = default
    def __missing__(self, key):
        return self._default      # <- no KeyError, but the zero value
class VulnerableReplica:
    """Real version: validity hangs on 'timestamp != 0'."""
    def __init__(self):
        self.messages  = EVMMapping(NULL_ROOT)   # message -> root
        self.confirmAt = EVMMapping(0)           # root    -> time
        self.confirmAt[NULL_ROOT] = 1            # the consequential initialization
    def is_message_proven(self, message):
        root = self.messages[message]            # stage 1: unknown => NULL_ROOT
        return self.confirmAt[root] != 0         # stage 2: NULL_ROOT => 1
class CanonicalReplica:
    """Canonicalization-guided version: the zero value is never
    a confirmed state, by invariant."""
    def __init__(self):
        self.messages  = EVMMapping(NULL_ROOT)
        self.confirmed = EVMMapping(False)       # default: not confirmed
    def confirm(self, root):
        if root == NULL_ROOT:
            raise ValueError("NULL_ROOT is not confirmable")
        self.confirmed[root] = True
    def is_message_proven(self, message):
        root = self.messages[message]
        if root == NULL_ROOT:                    # sentinel never valid
            return False
        return self.confirmed[root]
def test_i_break_unknown_message_counts_as_proven():
    """The I-break: one code (NULL_ROOT), two meanings."""
    assert VulnerableReplica().is_message_proven(UNKNOWN) is True   # vulnerable
def test_sentinel_separation_rejects_null_root():
    """The correction excludes the collapse."""
    canon = CanonicalReplica()
    assert canon.is_message_proven(UNKNOWN) is False
    try:
        canon.confirm(NULL_ROOT)
    except ValueError:
        pass
    else:
        raise AssertionError("NULL_ROOT should have been rejected")
def test_legitimate_root_still_works():
    """The correction does not break the legitimate function."""
    canon, known, root_a = CanonicalReplica(), b"\x11"*32, b"\xaa"*32
    canon.messages[known] = root_a
    assert canon.is_message_proven(known) is False    # not yet confirmed
    canon.confirm(root_a)
    assert canon.is_message_proven(known) is True     # now valid
\end{lstlisting}

The second check case is the actual evidence: it shows that a never-registered message counts as confirmed in the real version---because the same code \texttt{bytes32(0)} means both ``unknown'' and ``confirmed''.

\subsection{Complete EIP-712 Contract (Solidity)}

Compilable without error against \texttt{solc} 0.8.26 (optimizer runs = 200); \emph{compiled, not execution-tested}.

\begin{lstlisting}[language=Java]
// SPDX-License-Identifier: MIT
pragma solidity 0.8.26;
contract CanonicalEIP712Field {
    bytes32 public immutable DOMAIN_SEPARATOR;
    // keccak256("Order(string asset,uint256 amount,uint256 nonce)")
    bytes32 private constant ORDER_TYPEHASH =
        keccak256("Order(string asset,uint256 amount,uint256 nonce)");
    mapping(address => uint256) public nonces;
    error NonAsciiByte();
    error BadNonce(uint256 expected, uint256 got);
    error BadSigner();
    error MalleableSignature();
    error BadRecoveryId();
    // Upper bound of the low-S normal form for secp256k1 (n/2).
    uint256 private constant HALF_N =
        0x7FFFFFFFFFFFFFFFFFFFFFFFFFFFFFFF5D576E7357A4501DDFE92F46681B20A0;
    constructor(string memory name, string memory version) {
        DOMAIN_SEPARATOR = keccak256(abi.encode(
            keccak256(
              "EIP712Domain(string name,string version,uint256 chainId,address verifyingContract)"
            ),
            keccak256(bytes(name)),
            keccak256(bytes(version)),
            block.chainid,
            address(this)
        ));
    }
    // Canonical normal form of the case-insensitive field: ASCII lowercase.
    function canonicalizeAsset(string memory asset) public pure returns (string memory) {
        bytes memory b = bytes(asset);
        for (uint256 i = 0; i < b.length; i++) {
            uint8 c = uint8(b[i]);
            if (c > 0x7F) revert NonAsciiByte();
            if (c >= 0x41 && c <= 0x5A) b[i] = bytes1(c + 32);
        }
        return string(b);
    }
    // Structure hash over the CANONICALIZED field value.
    function hashOrder(string memory asset, uint256 amount, uint256 nonce)
        public pure returns (bytes32)
    {
        string memory canonAsset = canonicalizeAsset(asset);
        return keccak256(abi.encode(
            ORDER_TYPEHASH, keccak256(bytes(canonAsset)), amount, nonce
        ));
    }
    function digest(string memory asset, uint256 amount, uint256 nonce)
        public view returns (bytes32)
    {
        return keccak256(
            abi.encodePacked("\x19\x01", DOMAIN_SEPARATOR, hashOrder(asset, amount, nonce))
        );
    }
    // Verifies an order signature and consumes the per-signer counter.
    function verifyAndConsume(
        address signer, string memory asset, uint256 amount, uint256 nonce,
        uint8 v, bytes32 r, bytes32 s
    ) external returns (bool) {
        if (nonce != nonces[signer]) revert BadNonce(nonces[signer], nonce);
        // (1) Canonical signature form: the mirrored second form is rejected.
        if (uint256(s) > HALF_N) revert MalleableSignature();
        if (v != 27 && v != 28) revert BadRecoveryId();
        bytes32 d = digest(asset, amount, nonce);
        address recovered = ecrecover(d, v, r, s);
        // (2) Catch the zero-value fallback: address(0) is the error value.
        if (recovered == address(0) || recovered != signer) revert BadSigner();
        nonces[signer] = nonce + 1;
        return true;
    }
}
\end{lstlisting}

\section{Representation Divergence in Minimal Examples}

This appendix illustrates, at selected grid positions from Section~6.1, one minimal, concrete example each. It introduces no new statements and is not part of the contribution. Its purpose is purely didactic: where the main text develops the cases from real incidents, this appendix shows the underlying ambiguity at the byte or number level, small enough to verify by hand. The examples are deliberately simplified.

\subsection{Value Level, S-Break: the Two Valid s of an ECDSA Signature}

We work out the double assignment on a toy order $n = 23$. The example models only the modular $s$-mirroring of ECDSA verification, not a complete elliptic toy curve. Let the message hash $z = 17$, the private key $d = 6$, the one-time random number $k = 10$, and the value $r = 8$ obtained from the curve point be given. The signature component $s$ results as
\[ s = k^{-1} \cdot (z + r \cdot d) \bmod n. \]
With $k^{-1} = 7$ (since $10 \cdot 7 = 70 \equiv 1 \bmod 23$) it follows that
\[ s = 7 \cdot (17 + 8 \cdot 6) \bmod 23 = 7 \cdot 19 \bmod 23 = 133 \bmod 23 = 18. \]
The signature is thus $(r, s) = (8, 18)$. The form mirrored at the order is $n - s = 23 - 18 = 5$, thus $(r, n - s) = (8, 5)$. Both are valid, because the check uses $s$ only via its inverse and the two inverses stand negatively opposed to each other: $18^{-1} = 9$ and $5^{-1} = 14 = -9 \bmod 23$. The negated inverse mirrors the computed curve point at the $x$-axis; its $x$-coordinate, and thus the comparison value $r$, remains unchanged. Two number pairs, the same signature. The low-S normal form (Section~5.1) admits only $s \leq n/2$, here thus the 5, and removes the second form.

\subsection{Field-Value Level, S-Break: Several Encodings of the Same Value}

An amount field carries the value one hundred. Without a fixed normal form, several byte sequences are admissible that all decode as one hundred, such as the decimal strings ``100'' and ``0100'' and ``00100''. Semantically, all three denote the same value. As byte sequences they are different and therefore produce different hashes. If a replay protection hangs on the transaction hash, then the same payment counts as new in a second form (Section~6.2, Cosmos). Value canonicity (Section~7.3) requires here a fixed numeric normal form, such as without leading zeros, enforced before hashing.

\subsection{Composition Level, I-Break: Ambiguous Decomposition Without Separation}

Two fields are concatenated without a separator. The field values \texttt{"malle"} and \texttt{"ability"} concatenated yield \texttt{"malleability"}. But the same byte sequence also arises from \texttt{"mall"} and \texttt{"eability"}. The cut can lie before or after the \texttt{e}; without a length prefix or separator, the decomposition is not unique. Unlike in C.1 and C.2, here it is not one object present in several codes, but conversely: \emph{one} code \texttt{"malleability"} carries several valid field partitions and thus several meanings. This is the code-side direction (I-break), the non-injective packed encoding from the grid in Section~6.1. A signature formed over the concatenated form binds both meanings at once. This violates composition canonicity (Section~7.3). The correction is a uniquely decodable composition, such as a prepended length field per component, as canonical serializations such as BCS enforce (Section~6.3).

\subsection{State Level, I-Break: the Sentinel Collapse}

C.1 and C.2 show object-side multiple representation, multiple codes for one object; C.3 already switches to the opposite direction, one code with several meanings. This fourth case shows the same code-side direction at the state level, and it behaves fundamentally differently from the S-breaks.

The starting point is a peculiarity of the EVM storage model: if one queries a storage slot that was never written, it returns no error and no identifier for ``empty'', but silently the zero value \texttt{bytes32(0)}. An unset entry is thereby indistinguishable from one deliberately set to zero.

A contract now checks whether a message is already confirmed as valid, and reads for this a stored confirmation value. If, at initialization, precisely the zero value \texttt{bytes32(0)} is assigned the meaning ``confirmed'', then this one code carries two semantically different meanings:
\begin{quote}
\texttt{bytes32(0)} = ``no entry / unknown'' (because never set)\\
\texttt{bytes32(0)} = ``confirmed'' (because initialized so)
\end{quote}
The reading contract can no longer separate the two. An unknown, also a forged message leads to \texttt{bytes32(0)} and thereby counts as confirmed. Arbitrary messages pass through without proof (Section~4.2, Nomad).

The difference from C.1 to C.3 lies not in the break direction alone, but in the location of the failure. In C.1 and C.2, the ambiguity is removed by a value normal form before the binding, in C.3 by a uniquely decodable composition. Here, by contrast, the break sits in the reading step of the storage itself: the storage already returns an ambiguous value before any upstream normalization could engage. An upstream canonicalization does not apply, because the problem is not the representation of the input, but the assignment of code to meaning in the storage.

The correction is therefore not a normalization, but a redefinition of the assignment, sentinel separation (Section~5.2). A sentinel is a reserved special value for absence; the failure was that this absence value coincided with a valid state. One fixes by invariant that \texttt{bytes32(0)} may never mean ``confirmed'', and hangs validity on a separate, explicit marker whose default value for a never-set entry is ``not confirmed''. With this, absence and validity can no longer fall onto the same code.

\vspace{0.5em}
\noindent\emph{The four examples cover selected occupied grid positions from Section~6.1: object-side multiple representation (S-break) at the value and field-value levels, and code-side semantic collapse (I-break) at the composition and state levels. They do not replace the analysis of the real cases, but show their common root at the smallest possible instance in each case.}

\section*{Statement on the Use of AI Tools}

In the preparation of this work, AI language models were employed as tools, in particular Anthropic Claude Opus 4.8 and OpenAI GPT-5.5 Thinking. They served for linguistic revision, critical review of the argumentation, research and cross-checking of primary sources, the creation and execution of the appendix artifacts, and the verification of the worked-out examples. The responsibility for the content, all conceptual decisions, and the final version lies solely with the author.

\end{document}